\documentclass[12pt]{article}

\pdfoutput=1

\usepackage[margin=1.5in]{geometry}
\usepackage[comma,authoryear]{natbib}
\usepackage{geometry}
\usepackage{hyperref}
\usepackage{times}
\usepackage{amsmath}
\usepackage{amssymb}
\usepackage{amsthm}
\usepackage{verbatim}  
\usepackage{color}
\usepackage{graphicx}

\newcommand{\BRT}{$\mbox{{\rm BRT}}$}

\defcitealias{retrosheet}{Retrosheet}	
\defcitealias{baseballRef}{Baseball Reference}	

\newtheorem{definition}{Definition}

\begin{document}

\title{A Remark on Baserunning risk: Waiting Can Cost you the Game}

\author{Peter MacDonald \\
Department of Mathematics \& Statistics, McMaster University\\
	Hamilton, ON, Canada\\
        {\tt pdmmac@mcmaster.ca}\\
\and Dan McQuillan \\
Department of Mathematics, Norwich University\\
	Northfield, VT, USA\\
        {\tt dmcquill@norwich.edu}\\
\and Ian McQuillan\\
Department of Computer Science, University of Saskatchewan\\
	Saskatoon, SK, Canada\\
	{\tt mcquillan@cs.usask.ca}
}

\date{}

\maketitle

\begin{abstract}
We address the value of a baserunner at first base waiting to see if a ball in play falls in for a hit, before running. When a ball is hit in the air, the baserunner will usually wait, to gather additional information as to whether a ball will fall for a hit before deciding to run aggressively. 
This additional information 
guarantees that there will not be a double play and an ``unnecessary out''. However, waiting could potentially cost the runner the opportunity to reach third base, or even scoring on the play if the ball falls for a hit. This in turn affects the probability of scoring at least one run henceforth in the inning. 
We create a new statistic, the baserunning risk threshold (\BRT), which measures the minimum probability with which the baserunner should be sure that a ball in play will fall in for a hit, before running without waiting to see if the ball will be caught, with the goal of scoring at least one run in the inning. We measure a $0$-out and a $1$-out version of \BRT, both in aggregate, and also in high leverage situations, where scoring one run is particularly important. 
We show a drop in \BRT\ for pitchers who pitch in more high leverage innings, and a very low \BRT\ on average for ``elite closers''. 
It follows that baserunners should be frequently running without waiting, and getting thrown out in double plays regularly to maximize their chances of scoring at least one run.
\end{abstract}

\thispagestyle{empty}

\section{Introduction}

Consider the situation of a late inning of a close baseball game, with one out, and a single baserunner, who is on first base. Assume the batter hits the ball towards the outfield and it is not clear if the ball will fall in for a hit. The baserunner typically runs partway towards second base and then waits to see if the outfielder will catch the ball or not before continuing. We will refer to this strategy as the {\it conventional baserunning strategy}. This strategy is almost always used in order to prevent a double play. Indeed, if the ball is caught before it touches the ground, the baserunner must return to first base before the defense can get the ball to first base, or they are also ruled out, ending the inning. In this situation, standing and waiting can prevent a double play.

In contrast, when there are two outs, the baserunner has the freedom to run aggressively without waiting. Indeed, if the ball is caught, the inning is over anyway, so there is no possibility of a double play.

In this paper, we make the case that the 2 out, aggressive baserunning strategy should be employed with less than two outs in specific situations. In fact, this aggressive baserunning strategy is almost never used with fewer than two outs. In the rare cases when it is used, traditional baseball statistics cannot detect it, unless it results in a double play. Therefore, our arguments are necessarily indirect, rather than purely statistical.

Assume a runner is on first base in the 9th inning with one out, and a ball is hit to shallow right field. The outfielder is running very fast, attempting to catch the ball. Let us say that the probability of a hit is 0.5. Such a probability assessment could be made by the coaches, or the baserunner, in the time it takes the baserunner to reach the point where they would conventionally stop (we will discuss the problem of estimating this probability in Section \ref{conclusions}). In Section 2, we show that in this scenario, they should not stop. Indeed the only way stopping could be beneficial, is if the ball is actually caught. Therefore, the maximum reward of the conventional baserunning strategy is exactly one runner on first base, with two outs. We will show that this reward is relatively insignificant. As a clear example, consider that between the years of 2003 and 2008, late inning MLB specialist Mariano Rivera pitched in 58 different high leverage innings\footnote{We define {\em high leverage situations} to be either the eighth or ninth inning where the difference in score is at most one.} where there was exactly one baserunner, who was on first base with two outs. None of those 58 runners eventually scored in those innings. With hindsight, it is clear that no one should have coveted such a reward. Baserunners would have been better off pretending that there were two outs instead of one out, to increase their chances of scoring. Indeed, Rivera, is only slightly better than the average pitcher with a runner on third base and one out.

In the literature, there have been a number of similar questions addressed. In \citet{TheBook, mathletics}, the expected number of runs is shown, for each configuration of bases occupied, and by the event that occurs. The notion we are proposing is quite similar in nature to the notion of sacrifice bunting,  stealing bases, or aggressive baserunning (taking an ``extra'' base) which have been studied  (for example, also in \citealt{TheBook,BetweenNumbers,mathletics}). It is known, for example, that bunting reduces the expected runs in general, but can be a good idea for a poor hitter, or if scoring only one run is desired. 
In \citeauthor{BetweenNumbers}, the probability of scoring at least one run given each base configuration is given.
In general, runners should be far more aggressive in taking an ``extra base'' than they are in practice (as in \citeauthor{TheBook,mathletics}). In \citeauthor{TheBook}, the authors calculate the run value of taking an extra base. In this paper however, we are more concerned with the opportunities lost by waiting instead of running.

\section{Baserunning Risk Threshold}

Consider the possibility that there are $i$ outs where $i=0$ or $i=1$, and as a simplifying assumption, that aggressive baserunning without waiting, from first base, will result in the runner ending up on third base instead of second base should a ball in play fall in for a hit. We examine this reward --  the offense has a runner on third base instead of second base.  

\begin{definition}
We define the following, for $i \in \{0,1\}$:
\begin{itemize}
\item Let $T_i$ be the resulting probability of at least one run scoring later in the (half) inning, starting from a situation where there is a runner on third base and $i$ outs (no requirements regarding first and second base). 
\item Let $S_i$ be the probability of at least one run scoring later in the (half) inning, starting from a situation where there are $i$ outs and a runner on second base, but no runner on third base (no requirements regarding first base). 
\item Let $F_{i}$ be the probability that at least one runner will score later in the inning, starting from the situation that there are $i+1$ outs, a runner on first base and no runners on second or third.
\end{itemize}
\end{definition}

We will measure $T_i, S_i$ and $F_{i}$ for each pitcher, as well as in aggregate. To calculate $T_i$, as numerator, we use the number of half innings where there was a runner on third with $i$ outs and the pitcher pitching, and at least one run scores later on in the half inning. Therefore, even if the scenario occurs more than once in the same half inning, it is only counted once. As denominator we use the same value as the numerator plus the number of half innings where the pitcher was in that situation, and there were zero runs henceforth scored. We calculate $S_i$ and $F_{i}$ similarly.

 Note that the {\it reward} for the aggressive strategy with $i$ outs can be measured by the {\em difference} $T_i-S_i$, and that the only possible reward from the conventional strategy can be measured by $F_{i}$ (we use the subscript $i$ because $F_i$ is contributing to the $i$-out statistic, even though $F_i$ is calculated by examining situations where there are $i+1$ outs).

Now assume the ball is hit to the outfield and that the probability that the ball falls for a hit is judged to be $p$, where $0 \leq p \leq 1$. 
The probability that at least one run scores, using the conventional strategy is:
$$p \cdot S_i +(1-p)\cdot F_{i} .$$
The probability that at least one run scores with our proposed aggressive strategy is:
$$p \cdot T_i .$$
Thus, the aggressive strategy is at least as beneficial as the conventional strategy, with the goal of scoring a run in the inning, whenever:
\begin{equation} p \cdot T_i \geq p \cdot S_i+(1-p)\cdot F_{i}. \label{equation1} \end{equation}

Then, $BRT_i$ is the minimum value of $p$ for which risky baserunning is at least as good as the conventional strategy. In the unlikely event that $S_i \geq T_i$, then it is not possible that the conventional stradegy would be better than the aggressive strategy. (Observe that since $0 \leq p \leq 1$, Equation (\ref{equation1}) would have no solution). In this case, we automatically declare $BRT_i$ to equal $1$, meaning that one must be completely certain that the ball will fall in for a hit before running too far away from first base. 

If, on the other hand, $S_i<T_i$, then it makes sense to ask: which values of $p$ make the aggressive strategy at least as good or better? The answer is found by solving for $p$ in Equation (\ref{equation1}):
\begin{equation} p \geq \dfrac{F_{i}}{F_{i}+T_i-S_i}  \label{equation2} \end{equation}

We will therefore define \BRT\ as follows:
\begin{definition}
For $i \in \{0,1\}$ (the number of outs), we define the $i$-out {\em baserunning risk threshold}, ($\BRT_i$) as,
\begin{equation}
\mbox{{\rm BRT}}_i = 
\begin{cases}
\dfrac{F_{i}}{F_{i}+T_i-S_i} &   \mbox{if~} T_i-S_i>0,\\
1 & \mbox{if~} T_i-S_i \leq 0.
\end{cases}
\label{BRTdef}
\end{equation}
\label{BRTdefinition}
\end{definition}

Note that $T_i$ will almost certainly be bigger than $S_i$, unless our data is insufficient in the context used, to have confidence in those probabilities in the first place. If we did not include the second case in Equation (\ref{BRTdef}), then the \BRT\ statistic would be less than zero if and only if 
$S_i > T_i + F_{i}$, which would be undefined if and only if $S_i = T_i + F_{i}$, and would be
bigger than one if and only if $S_i > T_i$ and $S_i < T_i + F_{i}$. One of these scenarios, or $S_i = T_i$, is true if and only if $T_i - S_i \leq 0$, and therefore, the \BRT\ is mapped to one.

Definition \ref{BRTdefinition} was chosen based on two assumptions:
\begin{enumerate}
\item running aggressively without waiting to see if the ball in play is a hit, from first base, will result in the runner ending up safely at third base,
\item conventional baserunning would result in the baserunner ending up safely at second base.
\end{enumerate}
Our work shows that in order to maximize the probability of scoring at least one run under these assumptions, the aggressive baserunning strategy is better than conventional baserunning whenever $p\geq {\rm BRT}$. Without our assumptions the discussion would be much different.  In the event that running aggressively would result in the original baserunner being thrown out at third, our strategy would not be as effective. However, it is also possible that running aggressively could be much more effective as well, if for example, the runner from first could score if they run aggressively. This point will be discussed further in Section \ref{conclusions}.

\section{Aggregate and Pitcher BRT Statistics}
\label{statssection}

All data, unless noted otherwise, is from \citetalias{retrosheet}. We will give restrictions on the data used for the various statistics provided below.

The aggregate $\BRT_1$ and $\BRT_0$ statistics for all pitchers, in any inning from 1984-2011 appears in Table \ref{aggregatestat}.
\begin{table}[h]
\begin{center}{\small
\begin{tabular}{ l l }
$T_1= 0.627$ & $T_0 = 0.837$ \\
$S_1 = 0.398$ & $S_0 = 0.607$ \\
$F_1 = 0.142$ & $F_0 = 0.288$ \\ \hline
$\BRT_1 = 0.382$ & $\BRT_0 = 0.556$\\
\end{tabular} \hspace{.5in}
\begin{tabular}{ l l l }
$T_1= 0.624$ & $T_0 = 0.808$ \\
$S_1 = 0.398$ & $S_0 = 0.601$ \\
$F_1 = 0.133$ & $F_0 = 0.284$ \\ \hline
$\BRT_1 = 0.370$ & $\BRT_0 = 0.578$\\
\end{tabular} 
}
\end{center}
\caption{On the left, aggregate statistics for all pitchers in any inning from 1984-2011, for one and zero out. On the right is a restriction to high leverage situations.}
\label{aggregatestat}
\end{table}
This suggests that in any inning where the goal is to score at least one run, if there is one out, then the baserunner should run without waiting even if there is only a $38\%$ chance that a ball in play fall for a hit. This is quite clearly less than the conventional strategy (as discussed in Section \ref{conclusions}). Moreover, even if there is zero out, only a $56\%$ chance is required to make not waiting the best strategy. 
In addition, we calculate the aggregate statistics for all pitchers from 1984-2011, in {\em high leverage situations}, which we define to be either the eighth or ninth inning where the difference in score is at most one, also in Table \ref{aggregatestat}.


We next examine \BRT\ by partitioning pitchers based on the number of career high leverage innings they pitched in their career.  In all forthcoming statistics, we include career statistics in high leverage innings (ending in 2011) from all pitchers who either retired since 1984, or who are currently active. We include their career statistics even if they played before 1984, but retired after 1984. We summarize the data for $\BRT_1$ and $\BRT_0$ in
Table \ref{BRT1SummaryTable}.
\begin{table}[h]
\begin{center} {\small
\begin{tabular}{l|lllllllll}
$\mathbf{BRT_1}$ & 100 & 150 & 200 & 250 & 300 & 350+ & top 10 saves & all plays  \\ \hline
cumulative & 0.364 & 0.354 & 0.305 & 0.309 & 0.324 & 0.298 & 0.278 & 0.370   \\
mean & 0.425 & 0.411 & 0.338 & 0.329 & 0.344 & 0.317 & 0.302 & -  \\
standard dev & 0.260 & 0.237 & 0.176 & 0.137 & 0.117 & 0.119 & 0.111 & - \\
\end{tabular}
\begin{tabular}{l|llllllll}
$\mathbf{BRT_0}$ & 100 & 150 & 200 & 250 & 300 & 350+ & top 10 saves & all plays\\ \hline
cumulative & 0.569 & 0.576 & 0.576 & 0.520 & 0.518 & 0.545 & 0.528 & 0.578\\
mean & 0.623 & 0.619 & 0.645 & 0.590 & 0.555 & 0.574 & 0.540 & -  \\
standard dev & 0.263 & 0.250 & 0.230 & 0.254 & 0.194 & 0.137 & 0.107 & - \\
\end{tabular}
}
\end{center}
\caption{We provide the $\BRT_1$ (top) and $\BRT_0$ (bottom) for pitchers in high leverage innings.
We provide the statistics over all pitchers with career high leverage innings in the ranges
$[100,150), [150,200), [200,250), [250,300), [300,350)$ and $[350,\infty)$, 
as well as the top 10 all time career save leaders (save leaders
from \citetalias{baseballRef}), and all plays in high leverage situations. For each, we identify the
 cumulative statistics (without separating individual pitchers within each set), the mean statistics over all the
individual pitchers in each set, and the standard deviation between pitchers.}
\label{BRT1SummaryTable}
\end{table}

Table \ref{BRT1SummaryTable} demonstrates that, on average, the $\BRT_1$ is lower when examining sets of pitchers with fewer career high leverage innings pitched.
The cumulative statistic for the top 10 save leaders provides a $\BRT_1$
of $.278$, which is significantly less than the statistic of $.370$ for all pitchers in high leverage situations.
Hence, baserunners should be significantly more aggressive against elite closers.

For the zero out statistic, $\BRT_0$ also in Table \ref{BRT1SummaryTable}, we also see a tendency
for the $\BRT_0$ to be lower for pitchers with more high leverage innings, although there is a smaller difference in the cumulative $\BRT_0$ statistic between all pitchers, and top 10
save leaders ($0.578$ to $0.528$), than for the $\BRT_1$ statistic.

In Table \ref{All350PitchersBRT1}, we provide each pitcher's $T_1, S_1, F_1$ and $\BRT_1$ for all pitchers with 350 appearances in high leverage innings. We also collect the earned run average for each pitcher. The table is ranked in increasing order by $\BRT_1$.
\begin{table}[h]
\begin{center} {\footnotesize
\begin{tabular}{llllllllll}
Last Name & First Name & $T_1$ & $S_1$ & $F_1$ &  $\BRT_1$ & {\rm ERA} \\ \hline
Rivera & Mariano & 0.595 & 0.328 & 0.043 & 0.139 & 2.21 \\
Sutter & Bruce & 0.639 & 0.336 & 0.072  & 0.192 & 2.83 \\
Orosco & Jesse & 0.692 & 0.376 & 0.078 & 0.197 & 3.16 \\
Gossage & Rich & 0.658 & 0.354 & 0.078 & 0.204 & 3.57 \\
Righetti & Dave & 0.667 & 0.317 & 0.095 & 0.214 & 3.46 \\
Stanton & Mike & 0.707 & 0.359 & 0.099 & 0.221 & 3.92 \\
Fingers & Rollie & 0.638 & 0.338 & 0.094 & 0.239 & 2.90 \\
Minton & Greg & 0.543 & 0.330 & 0.068 & 0.243 & 3.10 \\
Jackson & Michael & 0.568 & 0.365 & 0.067 & 0.247 & 3.42 \\
Eckersley & Dennis & 0.667 & 0.338 & 0.117 & 0.262 & 3.50 \\
Hoffman & Trevor & 0.688 & 0.363 & 0.123 & 0.274 & 2.87 \\
Tekulve & Kent & 0.585 & 0.339 & 0.101 & 0.291 & 2.85 \\
McGraw & Tug & 0.638 & 0.436 & 0.084 & 0.295 & 3.14 \\
Jones & Doug & 0.694 & 0.468 & 0.094 & 0.295 & 3.30 \\
Franco & John & 0.671 & 0.330 & 0.143 & 0.295 & 2.62 \\
Smith & Lee & 0.603 & 0.283 & 0.136 & 0.298 & 3.03 \\
Reardon & Jeff & 0.549 & 0.356 & 0.082 & 0.298 & 3.16 \\
McDowell & Roger & 0.652 & 0.371 & 0.130 & 0.317 & 3.30 \\
Jones & Todd & 0.621 & 0.359 & 0.127 & 0.327 & 3.97 \\
Plesac & Dan & 0.591 & 0.385 & 0.118 & 0.363 & 3.64 \\
Timlin & Mike & 0.583 & 0.418 & 0.121 & 0.423 & 3.63 \\
Quisenberry & Dan & 0.543 & 0.392 & 0.115 & 0.432 & 2.76 \\
Hernandez & Roberto & 0.569 & 0.420 & 0.144 & 0.492 & 3.45 \\
Garber & Gene & 0.568 & 0.434 & 0.146 & 0.521 & 3.34 \\
Lavelle & Gary & 0.607 & 0.465 & 0.191 & 0.573 & 2.93 \\
Wagner & Billy & 0.436 & 0.375 & 0.084 & 0.580 & 2.31 \\ \hline
{\bf mean} &  & 0.614 & 0.371 & 0.106 & 0.317 & 3.17 \\
\end{tabular} }
\end{center}
\caption{The table above collects together all $\BRT_1$ data for all pitchers with 350 appearances in high leverage innings. Their corresponding $T_1, S_1, F_1$ contributing to their $\BRT_1$ is also
provided. Each pitcher's career earned run average is also given, from \citetalias{baseballRef}. The table is sorted in ascending order by $\BRT_1$.}
\label{All350PitchersBRT1}
\end{table}

\section{Discussion and Concluding Remarks}
\label{conclusions}

Many aspects of baseball strategy assume an ability to approximately judge probabilities in real time. For example, when the third base coach decides whether or not to send a runner home from third base on a sacrifice fly ball, the coach is (perhaps unwittingly) making a probability estimation that the runner can beat the throw to home plate, and comparing that to the probability the run will score in some other way. These estimated probabilities may not be accurately calculated, but they are likely approximately correct. They may be influenced by actual calculations and discussions before the game, and they may be adjusted after risks are taken and then reassessed. Therefore trial and experience helps improve accuracy. The conventional baserunning strategy, as defined in this paper, tacitly assumes that BRT is equal to 1. 

One other method that the baserunner or third base coach could use to estimate the probability that the ball fall in for a hit is to use {\it batting average on balls in play (BABIP)} in different contexts. If the ball is hit in some particular situation where the BABIP is greater than the \BRT, then the baserunner should run as soon as they determine that situation is occurring. For example, in \citet{babipbyangle}, the author calculates the BABIP depending on the horizontal angle of contact off of the bat (data from 2008). It is demonstrated that if the angle off the bat is between approximately $0$ degrees and $23$ degrees, then the BABIP is greater than $0.4$, which is greater than the aggregate $\BRT_1$. The estimate for the angle off of the bat can be made on contact, perhaps by the third base coach, and if it is less than this upper bound on the angle, then the runner should run immediately.


We could not tell from the statistical record, how often the aggressive baserunning strategy is employed. However, it seems to be extremely rare. This suggests that current practice of major league baseball teams is not approximately correct, in certain situations. By focusing on a shallow hit to right field, with the outfielder playing deep and running very fast, we have suggested a specific situation where it is very likely that current practice is not correct. There may be many other situations where current practice is not approximately correct, but it may be more difficult to make those determinations, until people start tracking that data necessary to make that analysis. Suppose a team starts to employ our suggested baserunning strategy. An observer would then keep track of the number of times that the runner does not stop. If a double play results, it counts as a failure. If the ball falls in for a hit, and the observer judges that the runner advanced further from running aggressively, then it is a success. With such data, much more substantial analysis would be possible. In other words, we do not know how to analyze our strategy directly, until a team tries it and records success.

\end{document}